\documentclass[useAMS,usegraphicx]{mn2e}
\usepackage{rotate}
\usepackage{times}
\newif\ifAMStwofonts
\AMStwofontstrue

\def\gs{\mathrel{\hbox{\rlap{\hbox{\lower4pt\hbox{$\sim$}}}\hbox{$>$}}}}
\def\ls{\mathrel{\hbox{\rlap{\hbox{\lower4pt\hbox{$\sim$}}}\hbox{$<$}}}}

\def\Msun{M$_{\odot}$}

\def\xmm{{\it XMM-Newton}}

\def\asca{{\it ASCA}}

\def\xmm{{\it XMM-Newton}}

\def\et{{et al.\ }}
\def\mcg{{MCG--6-30-15}}

\def\ngc{{NGC~4395}}

\def\erg{{\rm\thinspace erg}}

\def\Mpc{{\rm\thinspace Mpc}}
\def\Msun{\hbox{$\rm\thinspace M_{\odot}$}}

\def\s{{\rm\thinspace s}}
\def\ps{{\rm\thinspace s^{-1}}}

\def\ergps{\hbox{$\erg\s^{-1}\,$}}

\title[Exceptional variability of \ngc ]
      {The exceptional X-ray variability of the dwarf Seyfert
      nucleus \ngc}

\author[Vaughan \et]
       {S. Vaughan,$^{1}$
        K. Iwasawa,$^{2}$
        A. C. Fabian$^{2}$ and 
        K. Hayashida$^{3}$\\
$^{1}$X-Ray and Observational Astronomy Group, University of Leicester, Leicester, LE1 7RH\\
$^{2}$Institute of Astronomy, University of Cambridge, Madingley Road, Cambridge CB3 0HA\\
$^{3}$Department of Earth and Space Science, Osaka University, Osaka 560-0043, Japan
}
\date{accepted: 04/10/2004; submitted: 29/09/2004; in original form: 02/07/2004}
\pagerange{\pageref{firstpage}--\pageref{lastpage}}
\pubyear{2004}
\begin{document}
\maketitle
\label{firstpage}

\begin{abstract}
An analysis of the X-ray variability of the low luminosity Seyfert
nucleus \ngc, based on a long \xmm\ observation, is presented.  The
power spectrum shows a clear break from a flat  spectrum ($\alpha
\approx 1$) to a steeper spectrum ($\alpha \approx 2$) at a frequency
$f_{\rm br} = 0.5 - 3.0 \times 10^{-3}$~Hz, comparable to the highest
characteristic frequency found previously in a Seyfert galaxy.   This
extends the measured $M_{\rm BH}-f_{\rm br}$  values to lower  $M_{\rm
BH}$ than previous studies of  Seyfert galaxies, and is consistent
with  an inverse scaling of variability frequency with black hole
mass.  The variations observed are among the most violent seen in an
AGN to date, with the fractional  rms amplitude ($F_{\rm var}$)
exceeding $100$ per cent in the softest band.  The amplitude of the
variations seems intrinsically higher in \ngc\ than most other
Seyferts, even after accounting for the differences in characteristic
frequencies. The origin of this difference is not clear, but it is
unlikely to be a high accretion rate ($L/L_{\rm Edd} \ls 20$ per cent
for \ngc).  The variations clearly follow the linear rms-flux
relation, further supporting the idea that this is a ubiquitous
characteristics of accreting black holes.   The variations are highly
coherent between different energy bands with any frequency-dependent
time delay limited to $\ls 1$ per cent.
\end{abstract}



\section{Introduction}
\label{sect:intro}

The nucleus of the dwarf galaxy \ngc\ (Filippenko \& Sargent 1989)
lies at the edge of the parameters  space formed by Active Galactic Nuclei
(AGN). It is among the nearest ($D \sim 4.0~\Mpc$; Thim \et 2004)
and lowest luminosity ($L_{\rm Bol} \sim 5 \times 10^{40}~\erg~\ps$;
Moran \et 1999; Lira \et 1999) Seyfert galaxies known, and
harbours a central black hole with a  mass
probably in the range $M_{\rm BH} \sim 10^4 - 10^5 \Msun$ (Filippenko
\& Ho 2003), which compares to more luminous Seyfert galaxies with black
hole masses typically in the range $M_{\rm BH} \sim 10^6 - 10^8 \Msun$
(Wandel, Peterson \& Malkan 1999). In addition, it shows some of the
strongest X-ray variability of any
radio-quiet AGN (Iwasawa \et 2000; Shih, Iwasawa \&
Fabian 2003) with variations of nearly an order of
magnitude in a few thousand seconds. Yet in other respects, such 
as optical and ultraviolet emission (and absorption) line
properties, it resembles its more luminous counterparts (Filippenko \&
Sargent 1989; Crenshaw \et 2004).

The X-ray variations of Seyfert galaxies are characterised by a `red
noise' power spectrum (Lawrence \et 1987; M$^{\rm c}$Hardy
1989), approximated by a steep power law  at high frequencies
[$\mathcal{P}(f)\propto f^{-\alpha}$] with an index $\alpha \approx 2$
breaking to a flatter index $\approx 1$ below some break
frequency $f_{\rm br}$ (e.g. Edelson \& Nandra 1999; Nandra \& Papadakis 2001;
Uttley, M$^{\rm c}$Hardy \& Papadakis 2002; Vaughan, Fabian \& Nandra
2003; Markowitz \et 2003; M$^{\rm c}$Hardy \et 2004).  The prevailing
hypothesis is that $f_{\rm br}$ scales  inversely with the mass
of the black hole (or, equivalently, the break timescale scales with
$M_{\rm BH}$). Indeed, the limited data to date are consistent with a simple
scaling relation for all black holes, from Galactic
Black Holes (GBHs) in X-ray binaries ($M_{\rm BH} \sim 3-20~\Msun$; M$^{\rm
  c}$Clintock \& Remillard 2004) to
supermassive black holes in AGN ($M_{\rm BH} \sim 10^6 - 10^8
\Msun$). If this relation holds then the
break timescales in Seyferts may be used to infer the mass of the
central black hole by  comparison with the power spectra of well
studied GBHs such as Cygnus X-1 (Belloni \& Hasinger 1990; Nowak \et
1999) which contains a $M_{\rm BH} \approx 10 \Msun$ black hole 
(Herrero \et 1995). Uncovering the dependence of the characteristic variability
frequencies with the fundamental source parameters (mass, luminosity,
accretion rate, etc.) may help clarify the origin of the variability
itself. 

However, this simple hypothesis is difficult to test (see discussions in
Markowitz \et 2003; M$^{\rm c}$Hardy \et 2004 and Papadakis 2004). 
GBHs typically possess masses $M_{\rm BH} \sim 10 \Msun$ whereas
the majority of the well studied Seyfert galaxies have masses
$M_{\rm BH} \gs 10^6 \Msun$. Thus the $10^2 - 10^6 \Msun$
region of parameter space remains unexplored. \ngc, with its
unusually low mass black hole, falls in this critical region.
Previous studies of its X-ray variability (Shih \et 2003) using \asca\
suggested the expected power spectral break was present (at $f_{\rm br} \sim 3
\times 10^{-4}$~Hz), but this was poorly determined 
due to the interrupted sampling of the data (caused by the periodic
Earth occulations and SAA passages of the low Earth orbit). \xmm, with
its $\sim 2$ day orbit and high throughput, does not suffer from this problem
and has been providing the best constraints on the high frequency
power spectra  of Seyferts (Vaughan \et 2003a; Vaughan \& Fabian 2003;
M$^{\rm c}$Hardy \et 2004). The present paper describes a time series
analysis of the data from a long $\sim 100$-ks \xmm\ observation of
\ngc\ designed to confirm the break timescale and test the $M_{\rm
  BH}-f_{\rm br}$ scaling hypothesis. A future paper (Iwasawa \et in
prep.) will discuss other aspects of this observation.


\section{\xmm\ observation}
\label{sect:data}

\ngc\ was observed by \xmm\ (Jansen \et 2001) over 2003-November-30
and 2003-December-01 for a duration of $113,384$~s. This paper will
concentrate on the results from the EPIC pn camera which provided the
highest signal-to-noise X-ray data. 

Extraction of science products from the Observation Data Files (ODFs)
followed standard procedures using the \xmm\ Science Analysis System
v5.6.0 (SAS).  The data were taken with the EPIC
cameras in full-frame mode.
The field of \ngc\ contains several other bright
X-ray sources, although the nucleus itself is
not confused by these in the EPIC pn image. 
Source data were therefore extracted from the pn image using a circular region
(of radius $45$ arcsec) around the centroid of the nuclear
source.  Only events
corresponding to patterns $0-4$ (single and double pixel events) were
used for the pn analysis. Background events were extracted from
a region on the same CCD unaffected by source photons.  These
showed the background to be relatively low and stable throughout the
first $90$ ks of the observation.  During the final $\sim 15$~ks (as
the spacecraft approached the radiation belts at perigee) the
background rate increased dramatically; these data  were ignored for
the present analysis. The final extracted dataset contained $\approx
9.85 \times 10^4$ source counts. The source count rate was low enough 
that pile-up effects were negligible.

\begin{figure}
\centering
\includegraphics[width=6.40 cm, angle=270]{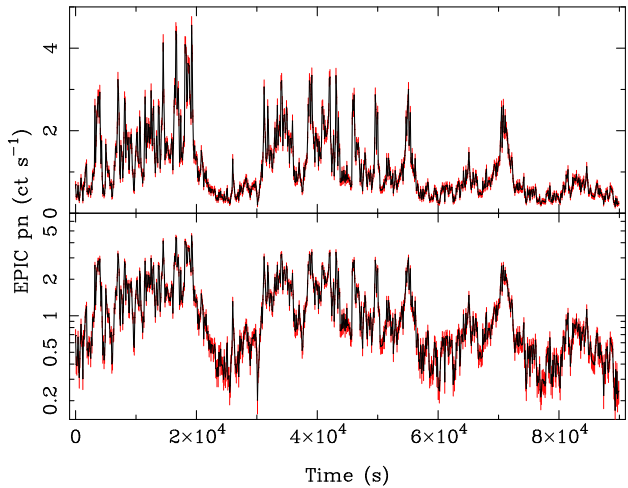}
\caption{
EPIC pn light curve of \ngc\ ($0.2-10$~keV with $100$~s bins).
The upper panel shows the light curve with the ordinate plotted on
a linear scale. The lower panel shows the same time series using a
logarithmic scale for the ordinate. 
\label{fig:lc}}
\end{figure}

Light curves were extracted from the EPIC pn data in four different
energy bands: $0.2-10.0$~keV (full band), $0.2-0.7$~keV (soft band),
$0.7-2.0$~keV (medium band) and $2.0-10.0$~keV (hard band). These were
corrected for telemetry drop outs (less than $1$ per cent of the total
time), background subtracted and binned to $50$~s time resolution. The
errors on the light curves were calculated by propagating the Poisson
noise.  The light curves were not corrected for the $\sim 71$ per cent
`live time' of the pn camera (which is only a
scaling factor).  The full band light curve is shown in
Fig.~\ref{fig:lc}. During the \xmm\ observation the average
$2-10$~keV flux was $\approx 5.6 \times 10^{-12}$~\ergps, similar to
the flux during the long \asca\ observation taken in 2001 ($\approx
4.7 \times 10^{-12}$~\ergps; Shih \et 2003), indicating the source was
in a fairly typical flux state.


\section{Time Series Analysis}
\label{sect:analysis}

The average source (background subtracted) and background count rates
are shown in Table~\ref{tab:basic} along with the fractional excess
rms variability amplitude of the source ($F_{\rm var}$; Vaughan \et
2003b) in each energy band.  The variability amplitude is remarkably
high, as can clearly be seen in the light curve (Fig.~\ref{fig:lc}). A
fractional rms in excess of unity is extremely rare and only occurs in
the most variable AGN known (NGC~4051, Green, M$^{\rm c}$Hardy \& Done
1999; IRAS~13224--3809, Boller \et 1997). Thus \ngc\ clearly deserves
to be compared to the most variable AGN. The
maximum-to-minimum variation through the length of the observation
is a factor $\sim 27$. 

\begin{table}
\centering
\caption{
Basic properties of the light curves. 
Mean source and background count rates and fractional excess rms
variability amplitudes ($F_{\rm var}$) in each of the four energy
bands (full $=0.2-10.0$~keV; soft $=0.2-0.7$~keV; medium $=0.7-2.0$~keV; 
hard $=2.0-10.0$~keV).
}
 \begin{center}
\begin{tabular}{lccc}                
\hline
Band  & source (ct s$^{-1}$) & background (ct s$^{-1}$) & $F_{\rm var}$ (\%) \\
\hline
full   & $1.11$ & $1.9\times 10^{-2}$  & $70.4\pm1.2$ \\
soft   & $0.28$ & $5.9 \times 10^{-3}$  & $105.9\pm1.9$ \\
medium & $0.34$  & $5.0\times 10^{-3}$  & $83.8\pm1.5$ \\
hard   & $0.49$  & $7.8\times 10^{-3}$  & $43.2\pm0.9$ \\
\hline
\end{tabular}
\end{center}
\label{tab:basic}
\end{table}


\begin{figure}
\centering
\includegraphics[width=6.0 cm, angle=270]{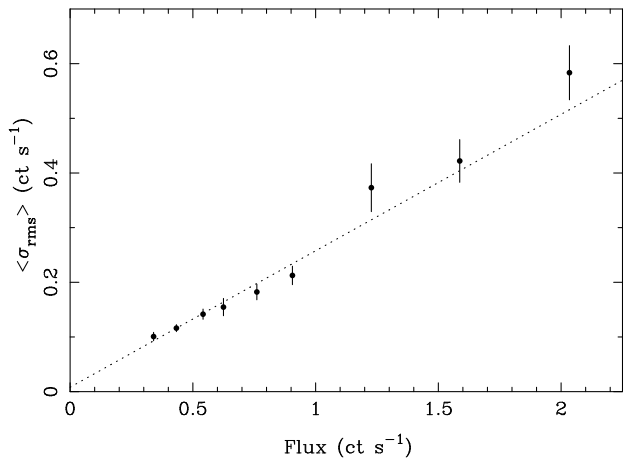}
\caption{
Rms-flux correlation. The flux and noise-subtracted rms were
calculated every $500$-s from the ($50$-s binned) full band light curve. 
These were then binned as a function of flux such that each bin
contains $\ge 20$ estimates. Clearly the mean rms is correlated
with the mean flux.
The dotted line shows the best-fitting linear function
($\chi^2 = 8.09 / 7 ~ dof$)
\label{fig:rms-flux}}
\end{figure}

\subsection{Rms-flux correlation}
\label{sect:rms-flux}

As can clearly be seen in the light curve, the distribution of fluxes
seems asymmetric, with a skewed tail of very high fluxes, interspersed
between periods of low and stable flux.  (As a crude test, examine the
light curve upside down.  It appears qualitatively different when viewed
this way indicating an asymmetry.)  One might be tempted to
naively interpret the high points as individual events (`flares'
or `shots') occurring between relatively quiescent periods. However,
there is a more  straightforward interpretation in terms of the
`rms-flux' relation known to exist in both Galactic X-ray binaries
(Uttley \& M$^{\rm c}$Hardy 2001) and Seyfert galaxies  (Vaughan \et
2003a; Vaughan \et 2003b; M$^{\rm c}$Hardy \et 2004).
Fig.~\ref{fig:rms-flux} demonstrates this relation operates in
\ngc. As discussed by Uttley, M$^{\rm c}$Hardy \& Vaughan (2004)
this is a natural artifact of a non-linear transformation of some
underlying linear variability process.  At high fluxes the rms
amplitude is proportionally higher, hence the light curve appears more
erratic, while at low fluxes the rms is lower and the light curve
smoother.  In effect the non-linear transformation is a `stretching' at
high fluxes and a `compression' at low fluxes.  The
effect on the light curve is more apparent in \ngc\ simply because of
its exceptionally large variability amplitude, which increases the
overall effect.  In order to illustrate this point the lower panel of
Fig.~\ref{fig:lc} shows the same light curve with a logarithmic flux
scale, which effectively transforms the distribution of data to a more
linear state and eliminates the linear rms-flux correlation. (On the
logarithmic scale the plot does not look so different from its upside
down counterpart.)  Thus the appearance of `flaring' and `quiescent'
periods does not require any non-linear dynamics but is simply a
consequence of the well-established rms-flux relation (which is a static 
non-linear effect).


\begin{figure}
\centering
\includegraphics[width=6.40 cm, angle=270]{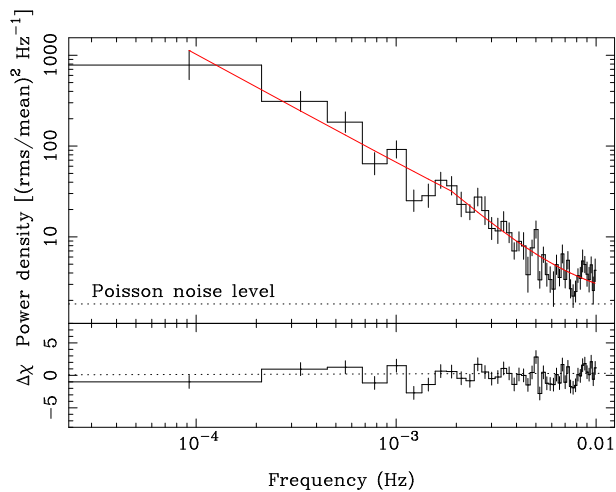}
\caption{
Log-binned periodogram. Upper panel: 
The histogram shows the binned
data ($N\ge 20$ per bin) with associated error bars. The dotted lines shows
the expected Poisson noise level. The solid curve marks the
best-fitting broken power law model (see text). 
Lower panel: residuals from the model fit.
\label{fig:psd}}
\end{figure}

\subsection{Power spectral analysis}
\label{sect:psd}

The power spectrum was examined by calculating the logarithmically
binned periodogram (following the prescription given in Papadakis \&
Lawrence 1993).  The result is shown in Fig.~\ref{fig:psd}, which
clearly reveals the red noise spectrum spanning at least two decades
of temporal frequency.

In order to compare the spectrum with those of other Seyferts, as well
as GBHs, and to determine the interesting power spectral parameters,
simple models were fitted to the binned periodogram.  The Poisson
noise in the light curve added a constant `background' level to the
spectrum, which was accounted for in the fitting by including a
constant at the appropriate level in the model (see Appendix A of
Vaughan \et 2003b). It is important to  note that this observation of
\ngc\ is free from spectral distortions such as `aliasing' and `red
noise leak' that commonly affect Seyfert  observations.  This was
demonstrated by comparison with Monte Carlo simulations (see below),
and implies that direct fitting of an analytical model to the binned
data is a perfectly reasonable procedure.

As a first test a simple power law was fitted to the binned
periodogram. This gave a rather poor fit ($\chi^2= 91.3 / 43 ~ dof$
for the full band data).  Allowing for a break in the power law gave a
substantial improvement ($\Delta \chi^2 = 14.9$ for $2$ additional
free parameters), significant at $\gs 97$ per cent in an
$F$-test.  The best-fitting broken power law model and the
data-model residuals are shown in Fig.~\ref{fig:psd}. The free
parameters were the power law index below the break ($\alpha_1$), the
index above the break ($\alpha_2$), the break frequency ($f_{\rm br}$)
and the normalisation.  Table~\ref{tab:psd} summarises the results of
fitting this model to the data from all four energy bands.
The low frequency slope is consistent with that obtained
from an analysis of the long \asca\ observation by Shih \et (2003;
$\alpha_1 = 0.98\pm0.18$),
but the break frequency is marginally inconsistent with this previous
analysis.  This may be partly due to the worse sampling (resulting in
some spectral distortion) for the \asca\ data, but could also indicate
the break frequency genuinely changed between the two observations.

\begin{figure}
\centering
\includegraphics[width=6.40 cm, angle=270]{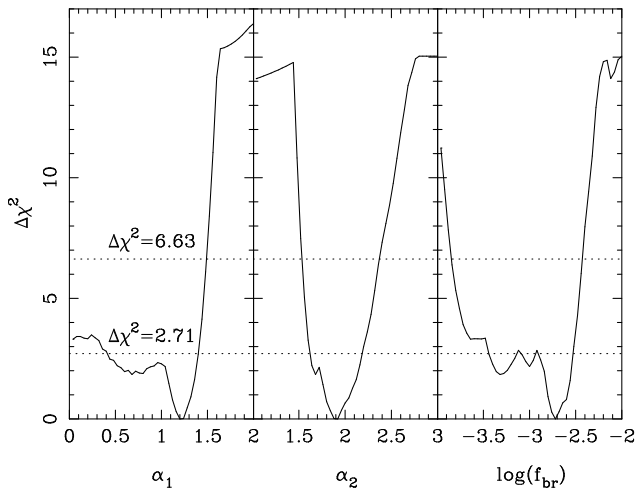}
\caption{
Parameter uncertainties. Variation in $\chi^2$ for the
broken power law model as the three parameters ($\alpha_1, \alpha_2$
and $f_{\rm br}$) are varied. The dotted lines show the $\Delta \chi^2
= 2.71, 6.63$ bounds, which delineate the $90$ and $99$ per cent
confidence bounds on the individual parameters.
\label{fig:fit}}
\end{figure}

The broken power law model provided a formally unacceptable fit to the
full band data, but the residuals (Fig.~\ref{fig:psd}) did not show
any obvious cause for alarm. The soft and medium band data were
reasonably well fitted using this model, while the  hard band remained
poorly fitted. Part of this may be due to the model having the wrong
shape break (e.g. too sharp), although fits with smoothly bending power
laws (see M$^{\rm c}$Hardy \et 2004) gave very similar quality fits
($\chi^2 = 76.39 / 41 ~ dof$ for the full band data).  Regardless of
whether  the shape of the break is more complicated than used in these
simple models, the existence of a break is clearly demonstrated. In
all the bands the best-fitting model breaks from $\alpha_1 \approx
1.2$ to $\alpha_2 \approx 2.0$ with the break frequency constrained to
lie in the range $f_{\rm br} \approx 0.5 - 3.0 \times 10^{-3}$~Hz.
Fig.~\ref{fig:fit} shows the change in $\chi_{\rm min}^2$ as each
parameter is varied.  Using the smoothly bending power law model
predicted a slightly lower break frequency ($f_{\rm br} =
0.8_{-0.6}^{+3.6} \times 10^{-3}$~Hz), compared to the sharply
breaking model,  together with a slightly flatter low frequency slope
$\alpha_1 \approx 0.7$.

\begin{table}
\centering
\caption{
Results of fitting the periodograms derived from the four energy
bands (full $=0.2-10.0$~keV; soft $=0.2-0.7$~keV; medium $=0.7-2.0$~keV; 
hard $=2.0-10.0$~keV)
with the broken power-law model (breaking from a slope of
$\alpha_1$ below $f_{\rm br}$ to $\alpha_2$ above). 
Errors on the model parameters correspond
to a $90$ per cent confidence level for one interesting parameter
(i.e. $\Delta \chi^{2}=2.71$).
\label{tab:psd}}
\begin{center}
\begin{tabular}{lcccc}                
\hline
          &                     &                     & $f_{\rm br}$ &               \\
Band      & $\alpha_1$          & $\alpha_2$          & ($10^{-3}$ Hz)           & $\chi^{2}/dof$  \\
\hline
full      & $1.20_{-0.76}^{+0.40}$ & $1.92_{-0.28}^{+0.24}$ & $1.9_{-1.5}^{+0.8}$ &  $76.4/41$   \\
soft      & $1.28_{-1.24}^{+0.16}$ & $2.08_{-0.44}^{+0.40}$ & $2.1_{-1.9}^{+0.7}$ &  $54.7/41$  \\
medium    & $1.20_{-0.12}^{+0.16}$ & $2.24_{-0.40}^{+0.36}$ & $3.0_{-0.5}^{+0.3}$ &  $59.6/41$   \\
hard      & $1.08_{-0.16}^{+0.12}$ & $2.32_{-0.48}^{+0.56}$ & $2.5_{-0.5}^{+0.6}$ &  $71.2/41$  \\
\hline
\end{tabular}
\end{center}
\end{table}

The Monte Carlo fitting procedure discussed by Vaughan \et (2003a; see
also Green \et 1999 and Uttley \et 2002) was also used to fit these
data. This takes into account any power spectral distortion due
to the time sampling of the data. However, in the case of \ngc\ there
were found to be no differences in best-fitting model
parameters between 
the simple (analytical) and robust (Monte Carlo) fitting techniques. 
There are two reasons for this. The first is that aliasing is
negligible due to the fact the data are contiguously binned from an
(effectively) continuous photon counting signal (van der Klis 1989). 
The second is that red noise leak was not present because the power
spectrum was flat $\alpha_1 \approx 1.2$ at the lowest measured
frequencies; leakage is only substantial when the low frequency
power spectrum remains steep ($\alpha_1 > 1.5$; Uttley \et 2002). 
Given that the binned periodogram is free from distortion, the simple 
fitting was used over the Monte Carlo method because it
is substantially faster and allowed for a much larger grid of model 
parameters to be computed in a reasonable amount of processing time.
(For most other Seyferts, where the power spectrum remains steep at the
lowest frequencies probed by \xmm, the Monte Carlo method should be used.)


\subsection{Cross-spectral analysis}
\label{sect:cross}

The cross-spectrum is closely related to the power spectrum.
It offers a comparison between two simultaneously sampled
time series (such as X-ray light curves in different energy bands) in
frequency space. The modulus of the cross-spectrum is used to define the
coherence (Vaughan \& Nowak 1997), a  measure of the
degree of correlation between variations in the two time series. 
If one dataset is simply a delayed or smoothed copy of the other, the
coherence will be unity. If the two datasets reflect totally
independent variations the coherence will be zero. The cross-spectrum
can also be used to measure the phase lag, or the time delay between
the two time series (assuming they are reasonable coherent). 
The coherence and phase lags of GBHs have been discussed by e.g. Nowak
\et (1999) and examined in Seyfert galaxies by Papadakis, Nandra \&
Kazanas (2001), Vaughan \et (2003a), Vaughan \& Fabian (2002) and 
M$^{\rm c}$Hardy \et (2004).

\begin{figure}
\centering
\includegraphics[width=6.40 cm, angle=270]{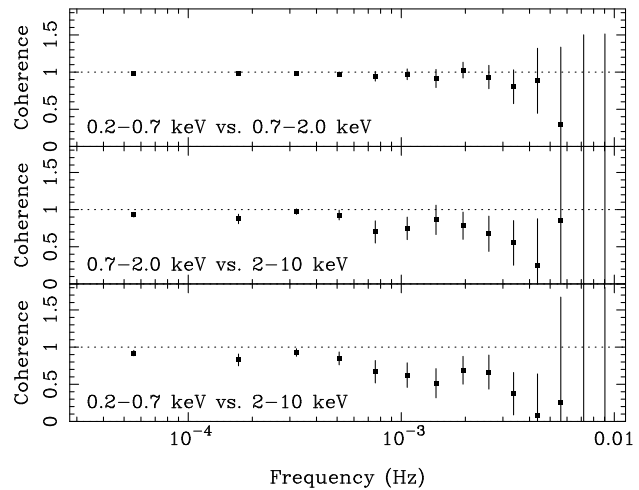}
\caption{
Coherence between energy bands.
The three panels show, for each combination of the three
energy bands, the coherence between the variations as
a function of frequency. The coherence
was calculated using the prescription of Vaughan \& Nowak (1997) and
binned into logarithmic frequency intervals 
ensuring that each bin contained $\ge 10$ cross-spectrum estimates.
Clearly at low frequencies the coherence is high $\gs 0.9$ in between
all the bands. At the highest frequencies the affect of Poisson noise
makes it difficult to measure the intrinsic coherence.
\label{fig:coh}}
\end{figure}

\begin{figure}
\centering
\includegraphics[width=6.40 cm, angle=270]{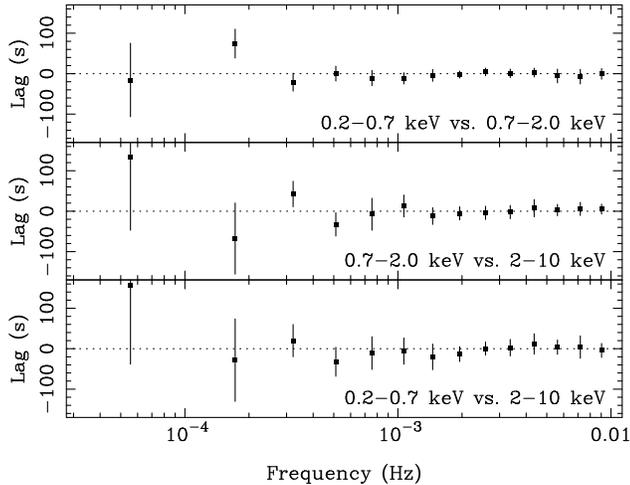}
\caption{
Time delay between energy bands.
The three panels show, for each combination of the three
energy bands, the time delay between the variations as
a function of frequency. The data were binned into logarithmic
frequency intervals  
ensuring each bin contained $\ge 10$ cross-spectrum estimates.
There are no clearly detected delays.
\label{fig:lag}}
\end{figure}

Fig.~\ref{fig:coh} shows the coherence between the three different
energy bands for \ngc. Clearly at low frequencies the variations are
highly coherent ($\approx   1.0$), while at high frequencies
the coherence becomes increasingly difficult to measure due to the
relatively stronger contribution from Poisson noise.  The coherence
was calculated using the method of Vaughan \& Nowak (1997) which
corrects for the effect of Poisson noise.  There is a slight tendency
for the coherence to fall off at higher  frequencies in the
soft-to-hard comparison (lower panel of the figure) which would be
consistent with that seen in other Seyfert 1 galaxies (e.g. Papadakis
\et 2001; Vaughan \et 2003a; M$^{\rm c}$Hardy \et 2004).
Fig.~\ref{fig:lag} shows the corresponding time lag spectra. It is
clear that there are no significant time delays between the bands,
even at the lower frequencies, with an upper limit of $\sim 100$~s at
$f \sim 10^{-4}$~Hz ($T \sim 10^4$~s), which corresponds to a $\ls 1$
per cent delay.  Unsurprisingly the cross-correlation functions
(computed using the Discrete Correlation Function of Edelson \& Krolik
1988) were reasonably symmetric and peaked at zero lag.

\section{Discussion}
\label{sect:disco}

\subsection{Summary of results}

The high frequency variability properties of \ngc\ have been examined
using a long ($\sim 100$~ks) \xmm\ observation.  The variations
clearly followed the linear rms-flux relation previously observed in
both GBHs and other Seyfert galaxies,  further supporting the idea
that this is a universal property of the X-ray variations from
accreting black holes.  The high frequency power spectrum was steep
$\alpha \approx 2$ above a break frequency $f_{\rm br} \approx 2
\times 10^{-3}$~Hz, below which it broke to a slope $\alpha \approx
1$. The break frequency is the second highest recorded to date for an
AGN (the record is presently held by Ark 564; Papadakis \et 2002). The
cross-spectrum between different energy bands shows the variations 
at frequencies $<f_{\rm br}$ were highly coherent and any frequency
dependent time delay is constrained to be shorter than $\tau \ls
0.01 f^{-1}$.  In other Seyferts, such as NGC~7469 (Papadakis \et 2001),
\mcg\ (Vaughan \et 2003a)  and NGC~4051 (M$^{\rm c}$Hardy \et 2004),
the delays are typically $\sim 1$ per cent. The limits of the time
delays for \ngc\ therefore did not rule out typical time delay
characteristics. Although the variations in these relatively
broad-band light curves were highly coherent, there is a strong
energy-dependence to the amplitude of the variations. This will 
be discussed in a forthcoming paper (Iwasawa \et in prep).

\subsection{The $M_{\rm BH} - f_{\rm br}$ relation}

\begin{figure}
\centering
\includegraphics[width=6.40 cm, angle=270]{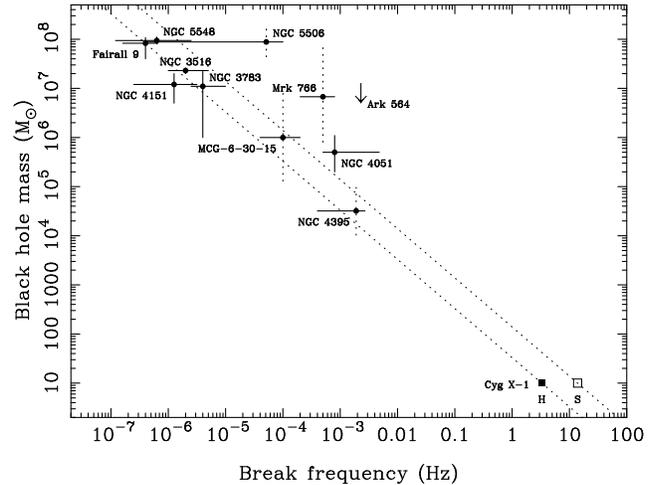}
\caption{
The $M_{\rm BH} - f_{\rm br}$ relation for $11$ Seyfert galaxies
including \ngc. 
The masses are from reverberation mapping experiments except
for the four objects marked using dotted error bars.
Also shown are typical
break frequencies for Cyg X-1 in both its low/hard and high/soft
states. The dotted lines show example $M_{\rm BH} \propto 1/f_{\rm br}$
relations consistent with the Cyg X-1 points.
\label{fig:m-f}}
\end{figure}

Fig.~\ref{fig:m-f} shows the $M_{\rm BH} - f_{\rm br}$ relation for
ten Seyfert galaxies using measurements from the literature (see also Markowitz
\et 2003; M$^{\rm c}$Hardy \et 2004; Papadakis 2004), together with
the new break frequency determined for \ngc.  The
power  spectral break frequencies from the other ten Seyferts were
collected from the following papers: Ark~564 (Papadakis \et 2002);
NGC~4051 (M$^{\rm c}$Hardy \et 2004); NGC~3516, NGC~3783, NGC~4151,
NGC~5548, Fairall~9 (Markowitz \et 2003); Mrk~766 (Vaughan \& Fabian
2003); \mcg\ (Vaughan \et 2003a) and NGC~5506 (Uttley \et 2002).
Black hole mass estimates were taken from the  reverberation mapping
studies where possible: Ark~564 (upper limit from Collier \et 2002);
NGC~4051 (Shemmer \et 2003); NGC~3516 (Wandel \et 1999); NGC~3783,
NGC~4151, NGC~5548, Fairall~9  (Kaspi \et 2000).  For NGC~5506 the
mass was estimated using the $M_{\rm BH}-\sigma_{\ast}$ relation
(section 2.3 of Papadakis 2004).  For Mrk~766 and \mcg\ the mass
estimates are very uncertain.  Based on its bulge luminosity, Uttley \et
(2002) argued for $M_{\rm BH} \sim 10^6$ in \mcg. Due to the
ambiguities in this method the error bars were taken to span an order
of magnitude in each direction.  
For Mrk~766, Wandel (2002) estimated $M_{\rm BH} \sim 10^7 ~ \Msun$
based on the optical luminosity and line widths, while Woo \& Urry
(2002) estimated $\sim 3.5 \times 10^6 ~ \Msun$ using the same
reasoning. For the purpose of comparison a value of $6.8 \times 10^{6}
~ \Msun$ (the mean of these two estimates) was used, again assuming an
order of magnitude uncertainty in either direction.  Also shown are
the typical frequencies of the high frequency break in Cyg X-1 in both
the low/hard and high/soft states.

For \ngc\ the break frequency $f_{\rm bf} = 1.9_{-1.5}^{+0.8}
\times 10^{-3}$~Hz was used together with a mass estimate of $M_{\rm
BH} = 10^4 - 10^5 ~ \Msun$. This is the `best guess' of Filippenko \&
Ho (2003) consistent with the available estimates: they
measured a firm upper limit on the mass of $\ls 6.2 \times 10^{6} ~
\Msun$  from stellar velocity measurements. Based on  photoionisation
modelling of optical emission lines, Kraemer \et (1999) derived an
estimate of $\sim 1.2 \times 10^5 ~ \Msun$, while Filippenko \& Ho
(2003) estimated $\sim 1.3 \times 10^4 ~ \Msun$ based on the
relationship between optical luminosity and broad line region size.
These are consistent with the mass estimated from the $M_{\rm
BH}-\sigma_{\ast}$ relation (Tremaine \et 2002): $\sim 6.6 \times 10^4
~ \Msun$. 
It is clear that, given the significant uncertainties in the mass
estimates, there is no glaring inconsistency between the new \ngc\
measurement and the 
hypothesised $M_{\rm BH} \propto f_{\rm br}^{-1}$ relation. However,
the details (such as whether the index of the relation is $-1$ and
which, if any, state of Cyg X-1 provides the best comparison) remain
to be investigated as more Seyfert power spectra are measured.

Hayashida \et (1998) estimated the  black
hole masses for Seyferts by directly comparing their X-ray power
spectra with that of Cyg X-1. They measured the frequency
$f_c$ at which the power spectrum reaches the level $f_c \times P(f_c)
= 10^{-3}$ and compared this to the corresponding frequency in Cyg
X-1. The motivation for using this frequency, instead of the break
frequency, is that the break is often harder to measure and (in
certain states at least) is known to change frequency in Cyg
X-1. There are a number of
underlying assumptions to this method.
The first is that both the slope and the (relative)
normalisation of the high frequency power spectra are similar for
Seyferts  and Cyg X-1, but the frequencies scale inversely with the
black hole mass. Implicit in this is the assumption that we are
using comparable energy bands in the two types of source (the power
spectrum is energy dependent). Furthermore, the original value of 
$f_c$ for Cyg X-1 used by Hayashida \et (1998) was taken from an
observation of 
the source in the low/hard state. This method therefore also assumes
that Seyferts most closely resemble GBHs in the low/hard state, which may
not be the case (Vaughan \et 2003a; M$^{\rm c}$Hardy \et 2004).
In the case of \ngc\ the frequency $f_c$ is not
directly observed; the power spectrum lies well above $f \times P(f) =
10^{-3}$ at all measured frequencies. Extrapolating the
best-fitting broken power law model to higher frequencies gives an
estimate of $f_c \sim 7.2\times 10^{-2}$~Hz (after accounting for the
factor of two difference in the normalisation used by Hayashida \et
1998). Scaling from the corresponding frequency in Cyg X-1, as defined
by equation 4 of Hayashida \et (1998) 
gives an estimate for $M_{\rm BH} \sim 6 \times 10^4 ~ \Msun$.
This is comparable to the estimates made by the other methods although if,
as discussed below, the normalisation of the power spectrum of \ngc\
is unusually high, this method will not be reliable.

\subsection{A universal power spectrum?}

The early work on the high frequency power spectra of Seyfert galaxies
found in favour of a universal power spectrum (Lawrence \& Papadakis
1993; Green \et 1993).  More recent work (Uttley \et 2002; Markowitz
\et 2003)  suggests that while most Seyferts exhibit a similarly steep
($\alpha \approx 2$) power spectrum at high frequencies, the slope
below the break may not be universal, and some objects may show a
second break (NGC~3783; Markowitz \et 2003) where others clearly do
not (NGC~4051; M$^{\rm c}$Hardy \et 2004).

The extraordinary amplitude of the variations in \ngc\ suggest that
the normalisation of the power spectrum may also vary quite
substantially from object to object. A fractional rms in excess of
$F_{\rm var} > 1$ has been noted in only two other Seyferts  prior to
\ngc\ (these are NGC~4051, Green \et 1999 and IRAS 13224--3809, Boller
\et 1997).  NGC~4051 is also thought to contain a relatively low mass
black hole (Shemmer \et 2003).  At first sight it remains possible
that the high variability amplitudes in both NGC~4051 and \ngc\ could
be due to their power spectra being shifted to higher frequencies
compared to Seyferts with more massive black holes (and, since the
spectra are red, more variance is shifted into the observable
frequency range) without the need for an intrinsically higher power
spectrum normalisation.  The power at the break multiplied by the
break frequency, $f_{\rm br} P_{\rm br}$, is indicative of the
total power in the spectrum.  For \ngc\ this is $\approx 5.4 \times
10^{-2}$ in units of $({\rm rms}/{\rm mean})^2$. The six Seyferts
studied by Markowitz \et (2003) have values  in the narrow range
$f_{\rm br} P_{\rm br} \approx 1.6 - 3.0 \times 10^{-2}$, all lower
than \ngc.  Papadakis (2004) estimated a `universal' value of $f_{\rm
br} P_{\rm br} \sim 1.7 \times 10^{-2}$ for Seyfert galaxies (called
${\rm PSD}_{\rm amp}$).  This means that, even after accounting for
the different regions of the power spectrum observed in different
objects, the normalisation of the  power spectrum of \ngc\ is
intrinsically high.

What could increase the overall amplitude of the variability (at all
frequencies)? A plausible suggestion is that an enhanced accretion
rate (relative to Eddington) somehow provides a more unstable
accretion flow and hence stronger fluctuations.  The accretion rate,
relative to the Eddington limit, can be estimated for \ngc\ using the
available black hole mass estimates, together with the bolometric
luminosity.  Moran \et (1999) and Lira \et (1999) estimate similar
values of $L_{\rm Bol} \sim 5 \times 10^{40}~\erg~\ps$.  Assuming a
mass in the range $M_{\rm BH} = 10^4 - 10^5 \Msun$ this gives an
Eddington fraction of $L/L_{\rm Edd} \sim 0.4 - 4$ per cent.  However, the mean
unabsorbed $2-10$~keV luminosity during the \xmm\ observation was
$L_{2-10} \sim 1.5 \times 10^{40}~\erg~\ps$.  Assuming $L_{\rm Bol} \gs
10 L_{2-10}$ would give $L_{\rm Bol} \gs 1.5 \times 10^{41}~\erg~\ps$
and  this, less conservative, luminosity estimate gives an Eddington
fraction of $L/L_{\rm Edd} \sim 1.2 - 12$ per cent\footnote{ This
assumes the Bolometric luminosity to be related to the mean X-ray
luminosity by $L_{\rm Bol} \gs 10 L_{2-10}$. If however the peak X-ray
luminosity was more representative of the overall  energy
distribution then the luminosity could be revised upwards by a further factor
of $\sim 3$.  }.  Thus \ngc\ is most likely accreting at a `normal'
rate; super-Eddington accretion
seems unlikely to be the origin of the unusually large
variability amplitude. Other possibilities include the black hole spin
parameter, $a$, and the inclination of the putative accretion
disc-black hole system, $i$. Unfortunately, these are more difficult
to test.


\section*{Acknowledgements}

Based on observations obtained with \xmm, an ESA science mission with
instruments and contributions directly funded by ESA Member States and
the USA (NASA). SV and KI thank the PPARC for financial support. 
ACF thanks the Royal Society for support.
We thank an anonymous referee for a useful report.

\bsp
\label{lastpage}

\end{document}